\documentclass[%
reprint,
superscriptaddress,
amsmath,amssymb,
aps,
prl,
floatfix,
]{revtex4-1}

\usepackage{graphicx}
\usepackage{dcolumn}
\usepackage{bm}
\usepackage{braket}
\usepackage[german,english]{babel}
\usepackage[colorlinks=true,linkcolor=blue,citecolor=blue,breaklinks]{hyperref}
\usepackage[pdftex]{color}
\usepackage{ulem}
\usepackage{comment}
\usepackage{natbib}
\usepackage{xcolor}

\usepackage{mathrsfs}

\begin{document}

\title{Fractal Quantum Phase Transitions: Critical Phenomena Beyond Renormalization}
\author{Zheng Zhou}
\affiliation{Department of Physics and State Key Laboratory of Surface Physics, Fudan University, Shanghai 200438, China} 
\author{Xue-Feng Zhang}
\affiliation{Department of Physics and Center of Quantum Materials and Devices, Chongqing University, Chongqing, 401331, China}
\author{Frank Pollmann}
\affiliation{Department of Physics, Technical University of Munich, 85748 Garching, Germany}
\affiliation{Munich Center for Quantum Science and Technology (MQCST), 80799 Munich, Germany}
\author{Yizhi You}
\affiliation{Princeton Center for Theoretical Science, Princeton University, NJ, 08544, USA}

\begin{abstract}
We identify a quantum critical point with fractal symmetry whose effective theory eludes the renormalization group framework. 
We consider the Newman-Moore model with three-body interaction subjected to an external transverse field, which exhibits a Kramers-Wannier type self-duality and a fractal $Z_2$ symmetry with Ising charge conserved on a fractal subset of sites, i.e., on Sierpinski gaskets.
Using large-scale quantum Monte Carlo simulations, we identify a continuous quantum phase transition between a phase with spontaneous fractal symmetry breaking and a paramagnetic phase.
This phase transition is characterized by the emergence of a fractal scaling dimension $d=\ln(3)/\ln(2)$ at the quantum critical point, where the power-law exponent of the correlation function is related to the fractal dimension of the
Sierpinski triangle. 
We develop a field theory to elucidate such quantum criticality and denote the fractal scaling as a subsequence of UV-IR mixing, where the low energy modes at the critical point are manipulated by short-wavelength physics due to the fractal symmetry.
\end{abstract}

\maketitle

\emph{Introduction}--- Quantum critical points connecting different phases have broad implications in modern quantum many-body physics. 
The universal features of critical phenomena are usually determined using the renormalization group~(RG) theory, emphasizing that the salient properties at the phase transition point can be understood by coarse-graining the local fluctuations and focusing on the physics at the long wavelength limit~\cite{sachdev2007quantum,sondhi1997continuous}. 
Based on this observation, a wide class of critical phenomena exhibits universal properties that are determined by symmetry, locality, and dimensionality~\cite{landau,wilson1983renormalization,fisher1983scaling,fisher1998renormalization}.
While the vast majority of known critical phenomena occur in presence of global symmetry, we investigate a special type of symmetry, dubbed \textit{fractal symmetry}~\cite{fracton_rev_2,subsys_symm_1,subsys_symm_2,subsys_symm_3}. 
Its symmetry charge is conserved on a subset of lattice sites whose volume scales with linear size $L$ as $L^d$ with some fractal dimension $d$ that is in general not an integer. 
Systems with fractal symmetries display interesting properties and are intimately related to fracton phases, an elusive state whose quasiparticles are immobile quasiparticles or with restricted motions on the subsystem~\cite{newman_1,newman_2,fracton_rev_1,fracton_rev_2,fracton_rev_3,type_i_1,type_i_2,type_i_3,type_ii_1,type_ii_2,subsys_symm_1,subsys_symm_2,subsys_symm_3,exact_solve_1,exact_solve_2,exact_solve_3,tensor_gauge_1,tensor_gauge_2,tensor_gauge_3,tensor_gauge_4,glass_1,glass_2,duality,fracton_rev_2,yoshida_2014,bulmash2018generalized,hsieh2017fractons,huang2018cage,aasen2020topological,gromov2020fracton,gromov2019towards,devakul2020fractalizing,williamson2020type,dua2019sorting,dua2019compactifying}. 
The current revival of critical phenomenon enriched by various symmetries motivates several outstanding questions: Can we find quantum critical points triggering the phase transition with spontaneous fractal symmetry breaking? What is the impact of fractal symmetry in these critical phenomena and how are they distinct from the ones that occur in presence of global symmetries~\cite{you2020higher,you2020fracton,seiberg2020exotic,xu2007bond,paramekanti2002ring,tay2011possible,xu2008resonating,karch2020reduced,gorantla2021modified}?

\begin{figure}[t]
    \centering
    \includegraphics[width=\linewidth]{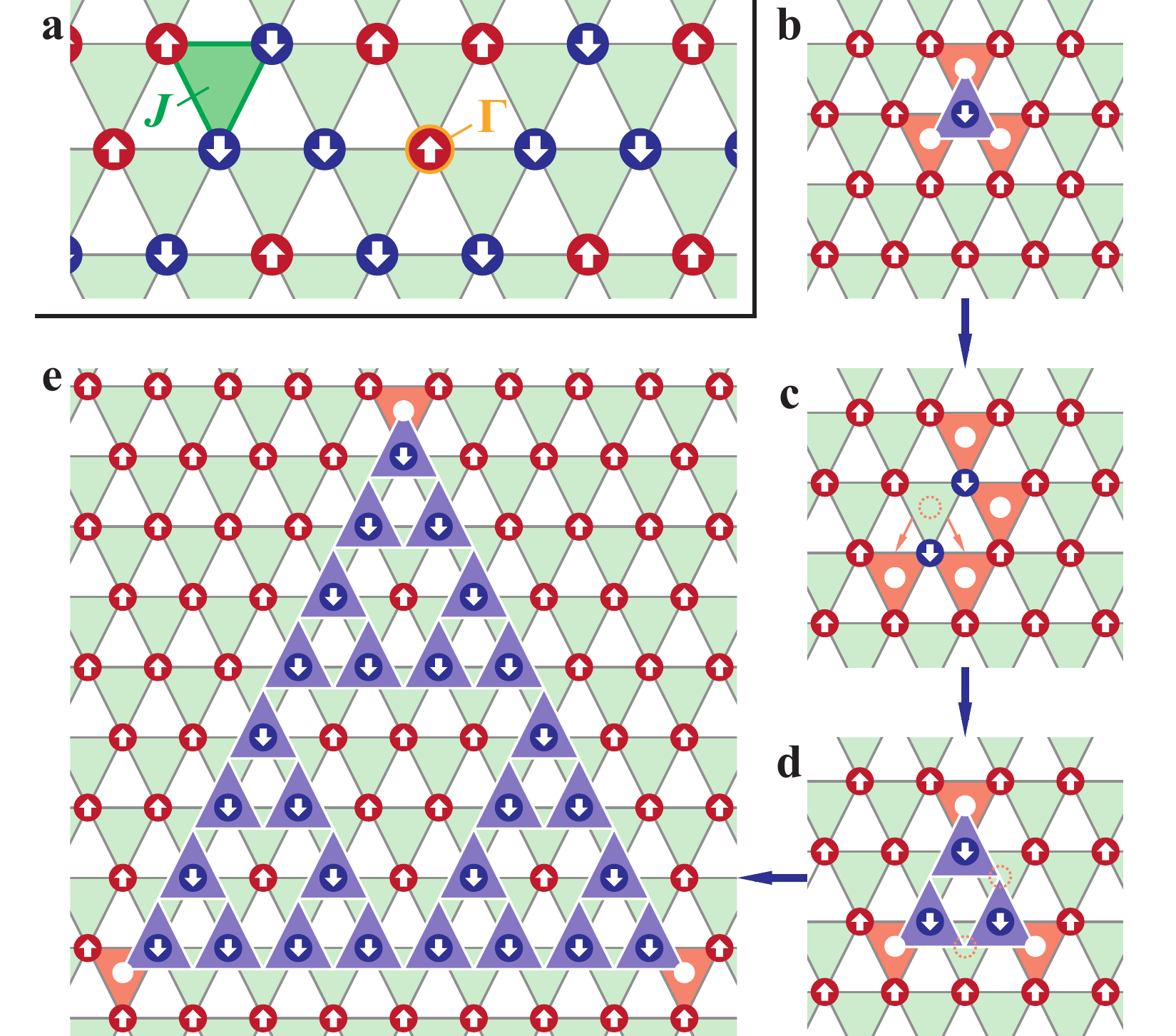}
    \caption{ Illustration of the Newman-Moore model setup and its fracton excitations. (a) The model Hamiltonian and a configuration in the ground state manifold. (b--d) The creation and separation of three fracton excitation (red triangle) from the ground state by flipping the spin inside the Sierpinski gaskets (blue). The fracton lives at the corner of a Sierpinski triangle. (e) The motion of fractons from the configuration of (b). Separating the three fractons living at the corner of the fractal requires the growth of the Sierpinski triangle, with intermediate steps engendering additional excitations.}
    \label{fig_1}
\end{figure}

In this work, we address these questions and propose a new type of quantum critical point, characterizing the phase transition with spontaneously fractal symmetry breaking. 
Our starting point is the 2D Newman-Moore model~\cite{newman_1,newman_2} on a triangular lattice, consisting of three-body Ising couplings on each down-pointing triangle---which was initially proposed to study glassiness at finite temperature. 
This model accommodates a remarkable fractal symmetry where the Ising charge is not conserved globally but on a subset of sites reminiscent of the Sierpinski gaskets, whose volume scales with linear size $L$ as $L^{\ln(3)/\ln(2)}$. 
In the presence of a transverse field, the model exhibits Kramers-Wannier type self-duality~\cite{fracton_rev_2,duality} with a phase transition at the self-dual point connecting the paramagnetic phase and fractal symmetry breaking phase. 
The fractal symmetry breaking phase has a subextensive degeneracy sensitive to the system size~\cite{newman_2,newman_1,fracton_rev_2}. 
Isolated excitations above the ground state are \textit{fractons} that cannot be moved individually and the spatial separation of such isolated excitations requires one to overcome a logarithmically growing energy barrier~\cite{newman_1,newman_2}.

We carry out large-scale quantum Monte Carlo (QMC) simulations using the stochastic series expansion (SSE) algorithm~\cite{qmc_1,qmc_2,qmc_3,sm} to investigate the fractal symmetry breaking phase transition in the Newman-Moore model with transverse field.
We use periodic boundary conditions in both spatial directions and perform finite-size scaling while keeping the inverse temperature $\beta$ proportional to $L$ to extrapolate the thermodynamic limit at zero temperature. 
The glassiness of the Newman-Moore model~\cite{newman_2,glass_1,sm} makes the simulations challenging in the ordered phase due to local minima with very close energies.
To facilitate the inter-minima transition, we use an optimized update scheme for the three-body Ising operator.
We detect the fractal symmetry breaking through a many-body correlator~\cite{fracton_rev_2} and find indications for a continuous phase transition. 
In particular, we observe that the power-law correlations at the phase transition point exhibit an emergent fractal scaling dimension $\langle\sigma^x(0)\sigma^x(r)\rangle=r^{-2\ln(3)/\ln(2)}$, which exactly matches the dimensionality of the fractal Sierpinski triangle $d=\ln(3)/\ln(2)$. 
We then develop an effective field theory to characterize the critical point and justify the elusive fractal scaling dimension.
Peculiarly, this quantum critical point is a characteristic example of ``UV-IR'' mixing~\cite{gorantla2021modified,xu2007bond,paramekanti2002ring,you2019emergent,you2020fracton,you2021fractonic} that escapes from the conventional renormalization group perspective. 
As opposed to the RG paradigm in critical theory, where the low energy modes are controlled by long wave-length physics, the fractal symmetry at the critical point engenders a subextensive number of short wave-length modes with low energy. 
Hence, the quantum dynamics and universal behavior at the phase transition point are dominated by the UV modes with strong local fluctuations. 
The fractal scaling dimension is a direct subsequence of such UV-IR mixing with excessive low energy fluctuations at a short wave-length scale.

\emph{Model}---
We consider the quantum Newman-Moore model on a triangular lattice~\cite{newman_1,newman_2} in presence of a transverse field,
\begin{equation}
    H=-J\sum_{(ijk)_\triangledown}\sigma_i^z\sigma_j^z\sigma_k^z-\Gamma\sum_i\sigma_i^x,
    \label{model}
\end{equation}
where we set $J = 1$ and vary $\Gamma>0$. 
The first term is a three-body Ising interaction on each downward triangle~(green) as shown in Fig.~\ref{fig_1}. 
Note that the three-body interactions break the $\mathbb Z_2$ Ising symmetry and the $C_6$ lattice symmetry. 
In the large transverse field limit $\Gamma\gg 1$, the system is in the paramagnetic phase with spins polarized along $x$ direction. 
In the small transverse field limit $\Gamma\ll 1$, the ground state exhibits patterns in which downward triangles have an even number of $|\downarrow\rangle$ spins~\cite{newman_1}. 
The elementary excitation is a downward triangle with an odd number of $|\downarrow\rangle$ as illustrated in Fig.~\ref{fig_1}. 
These excitations, dubbed ``fractons'', contain interesting dynamical features, including restricted mobility that a single excitation cannot move alone. 
By flipping an individual spin from the ground state, we simultaneously create three fracton excitations at the three adjacent downward triangles. 
A further separation of the three fractons with distance $2^L$ is accomplished by flipping the spins on Sierpinski gaskets.
As a result, the fracton excitations live on the corners of a Sierpinski triangle and the growth of the Sierpinski triangle requires intermediate steps that engender additional excitations with an energy barrier $E(r)\sim \ln(r)$ for a distance $r$~\cite{newman_2}. 

While the three-body breaks the global $\mathbb Z_2$ symmetry, the model displays an exotic fractal symmetry generated by a Cellular-Automaton~\cite{fracton_rev_2,sm}. 
In particular, a symmetry operator on a Sierpinski gasket $\prod_{i\in\mathrm{ST}}\sigma^x_i$ commutes with the three-body interaction in Eq.~(\ref{model})~(up to corner terms).
Thus the Hamiltonian displays a \emph{fractal} $\mathbb Z_2$ symmetry with the Ising charge conserved on the subextensive manifold on Sierpinski triangles, as shown in Fig.~\ref{fig_1}e, of fractional dimensionality $d=\ln(3)/\ln(2)=1.58$. 

We will make use of the fact that Hamiltonian Eq.~(\ref{model}) has a Kramers-Wannier self duality~\cite{fracton_rev_2,duality,sm}. 
In the dual picture, the Ising degree of freedom $\tau^x$ living at the center of each down-pointing triangle is dual to the three-body interaction $\sigma_i^z\sigma_j^z\sigma_k^z$ in the original model. 
Likewise, the transverse field $\sigma_i^x$ in the original Hamiltonian is mapped to a three-body interaction $\tau_i^z\tau_j^z\tau_k^z$ on upward triangles of the dual lattice. 
Thus, the dual Hamiltonian is reminiscent of the original one by swapping $\Gamma$ and $J$.

\begin{figure}[t]
    \centering
    \includegraphics[width=1\linewidth]{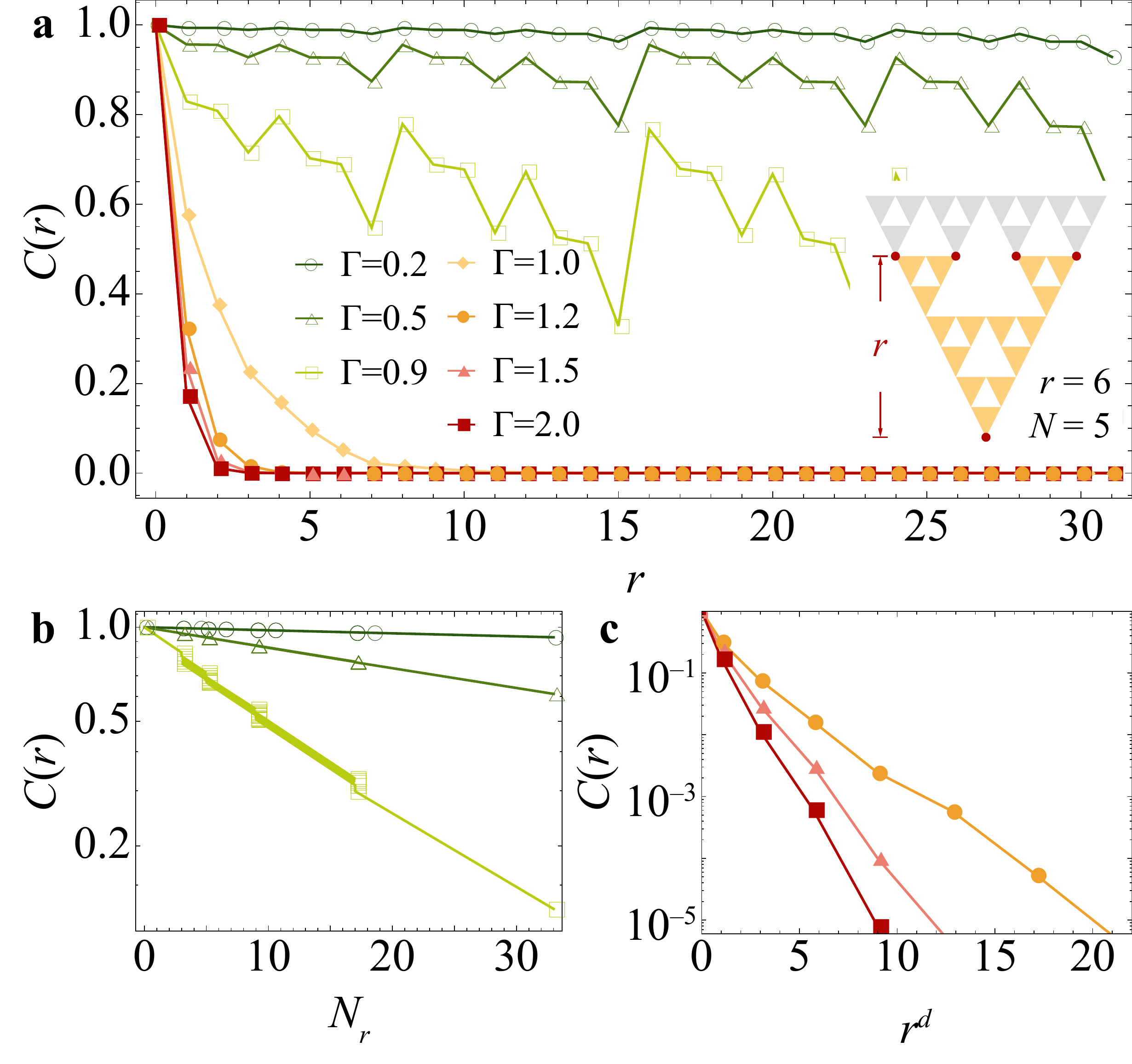}
    \caption{ Many-body correlation function $C(r)$ distinguishing the fractal ordered phase and the paramagnetic phase. (a) Distinct behaviors of $C(r)$ at different transverse field $\Gamma$. Inset: An illustration of $C(r)$ as the product of $\sigma^z$ at the marked sites at $r=6$ and $N=5$ (b) The correlator $C(r)$ scales exponentially with number $N_r$ of $\sigma^z$ in the product in the ordered phase and (c) decays exponentially with the streched distance $r^d$ in the paramagnetic phase.}
    \label{fig_2}
\end{figure}

\textit{Many-body correlators}--- 
It was demonstrated in Refs.~\onlinecite{newman_1,newman_2} that the ground state of the three-body term breaks discrete scale invariance and has a degeneracy that grows subextensively with the system size in a non-monotonic way. 
Particularly, the ground state is unique for system sizes $2^N \times 2^N $ and maximally degenerate when the size is $(2^N-1) \times (2^N-1)$ for an integer $N$. 
The ground state manifold consists of all constraint-satisfying patterns with an even number of spin-down per downward triangle, and each classical pattern in this manifold spontaneously breaks the $\mathbb Z_2$ charge conservation on the Sierpinski fractal. 
Such peculiar symmetry breaking pattern can be characterized by a many-body fractal correlator~\cite{fracton_rev_2,sm}. 
For this we first define a translation operator $\mathscr T$ as $\mathscr T(\sigma_{ij})=\sigma_{i,j-1}\sigma_{i-1,j-1}$. 
By acting this operator repeatedly as $\mathscr T^r(\sigma_{ij})=\mathscr T(\mathscr T^{r-1}(\sigma_{ij}))$, we obtain a series of product of the Pauli operator $\Pi_r=\mathscr T^r(\sigma_{00})$ along the fractal structure. %
The correlation function of the $\Pi$ operator is defined as
\begin{equation}
    C(r)=\langle\Pi_0\Pi_r\rangle=\langle\sigma_{00}^z\mathscr T^r(\sigma_{00}^z)\rangle.
    \label{eq_cor}
\end{equation}
This many-body correlator has a non-zero expectation value in the fractal symmetry breaking phase while it goes to zero for $r\rightarrow\infty$ in the paramagnetic phase. 
When $r=2^n$, the correlation function reduces to the three-body correlation between the three spins living at the corner of the fractal operator $C(r)=\langle\sigma^z_{0,0}\sigma^z_{2^n,0}\sigma^z_{2^n, 2^n}\rangle$. 
The nonvanishing expectation value of such fractal operator implies long-range correlations between the spins living at the corner of the Sierpinski triangle. 
An alternative way to visualize this many-body correlator is from the dual picture:
As the operator in Eq.~(\ref{eq_cor}) can be obtained by the product of all $\tau_x=\sigma_i^z\sigma_j^z\sigma_k^z$ operator inside the Sierpinski fractal, the many-body operator counts the dual-charge of $\tau_x$ inside the fractal geometry. 
In the limit $\Gamma\to0$, all ground state satisfy $\tau_x=\sigma_i^z\sigma_j^z\sigma_k^z=1$ such that the many-body correlator converges to unity. 
The transverse field flips a sparse number of spins from the ground state. 
Each spin-flip creates three fracton defects with $\tau_x=\sigma_i^z\sigma_j^z\sigma_k^z=-1$ in the adjacent downward triangle. 
In the fractal symmetry breaking phase, an isolated fracton defect is confined and the three fractons bound state can only change the dual-charge inside the Sierpinski triangle provided they live at the corner of the fractal. 
Thus, the scaling of such a many-body operator obeys a ``corner-law'' which depends on the number of operators in Eq.~(\ref{eq_cor})~\cite{sm}. 

We confirm this scaling behavior in our Monte-Carlo simulations shown in Fig.~\ref{fig_2}.
In the ordered phase with $\Gamma<1$, the many-body correlation displays long-range order as a manifestation of the fractal symmetry breaking nature.
Notably, an oscillation behavior is observed, which confirms the breaking of discrete invariance. 
The argument above dictates that $C(r)$ depends on the number $N_r$ of $\sigma_z$ in the product. 
Numerically, we observe that at small $\Gamma$, data with the same $N_r$ but different $r$ collapse well, whereas this collapse begins to fail for the system sizes considered when approaching the critical point (e.g., $\Gamma=0.9$), as the correlation length increases and surpasses the magnitude of system size. 
In the paramagnet phase with $\Gamma>1$, the many-body correlator shows a stretched decaying behavior $C(r)\sim\exp(-\alpha r^d)$, with $d$ being the dimension of the Sierpinski triangle. 
The strong transverse field proliferates the fracton defects so the isolated defects are spatially separated at a larger distance. 
As the defect inside the Sierpinski fractal flips the dual-charge and suppresses the many-body correlator, the exponential suppression of the many-body operator should scale with the Sierpinski triangle area $S=r^d$. 

\begin{figure}[t]
    \centering
    \includegraphics[width=1\linewidth]{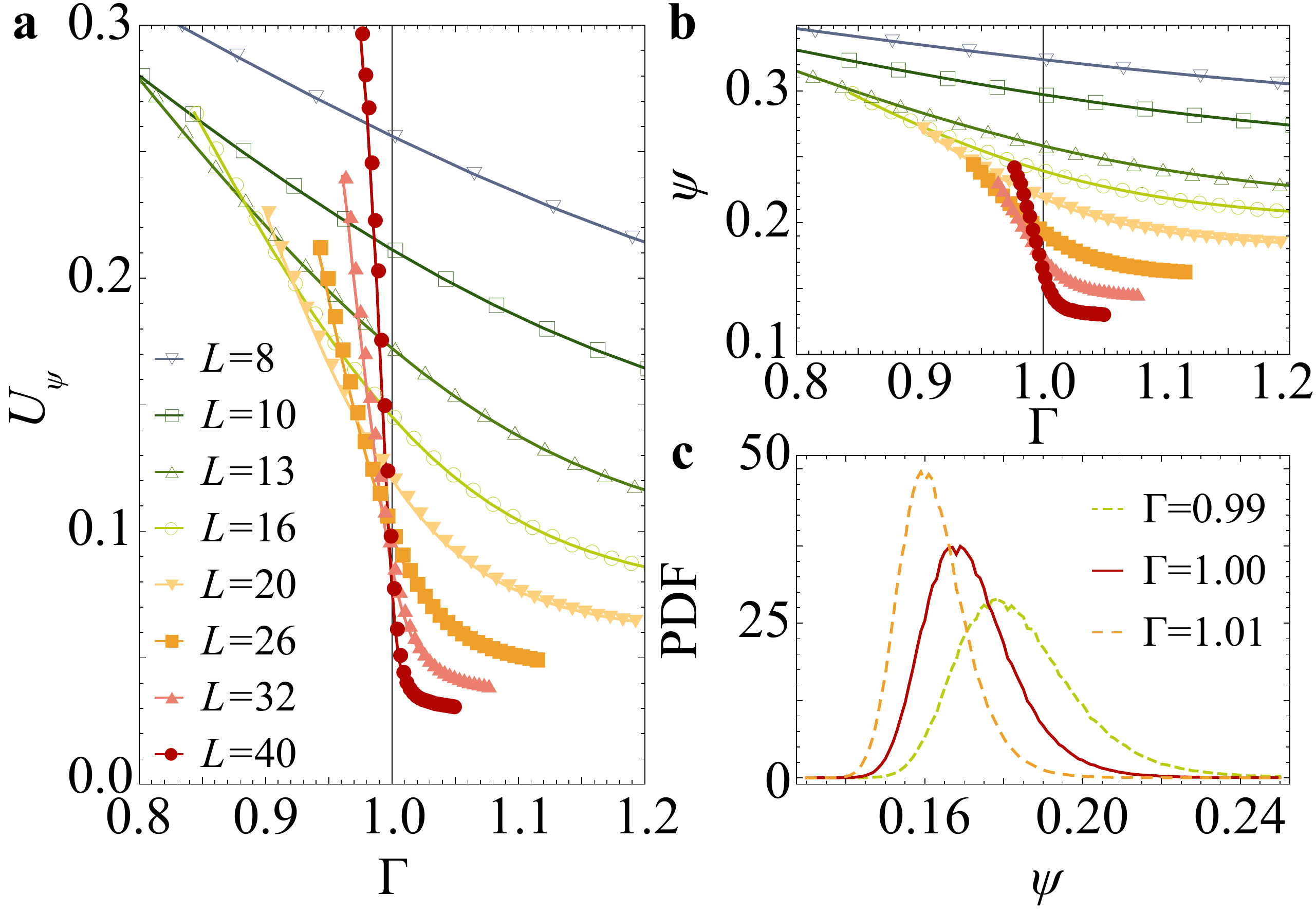}
    \caption{The order parameter $\psi$ as an evidence of continuous phase transition. (a) The Binder cumulant of the order parameter $U_\psi$ and (b) the order parameter $\psi$ itself as a function of the transverse field $\Gamma$ with different system sizes $L$. The black vertical line denotes the position of critical point. (c) The probability density function (PDF) of the order parameter $\psi$ in vicinity of the critical point, measured with system size $L=32$. }
    \label{fig_3}
\end{figure}
\textit{Order parameter} --- The phase transition can also be characterized by an order parameter defined as the summation of all the fractal $\Pi$ operators,
\begin{equation}
    \psi=\frac{1}{L}\left|\sum_{i=0}^{L-1}\Pi_i\right|.
    \label{eq_order}
\end{equation}
This quantity has a non-zero expectation value in the ordered phase as its square can be rewritten as the sum of all the correlators,
\begin{equation}
    \langle\psi^2\rangle=\frac{1}{L^2}\sum_{i,j=0}^{L-1}\langle\Pi_i\Pi_j\rangle.
\end{equation}
We measure the order parameter $\psi$ and its Binder cumulant $U_\psi=\left(3-\langle\psi^4\rangle/\langle\psi^2\rangle^2\right)/2$ in the Monte Carlo simulation and choose system sizes for which the ground states are unique. 
As shown in Fig.~\ref{fig_3}, the order parameter $\psi$ scales to a finite value when $L\to\infty$ in the ordered phase and scales to zero in the paramagnetic phase. 
For sufficiently large system sizes, the Binder ratio curves intersect at $\Gamma_c=1$, which points towards a continuous phase transition. 
Another strong indication for a continuous phase transition stems from the histogram of the order parameter. 
Instead of having a double peak structure, which would be typical for a first order transition, we find that the peak in probability density function moves continuously when crossing the critical point. 

\begin{figure}[t]
    \centering
    \includegraphics[width=1\linewidth]{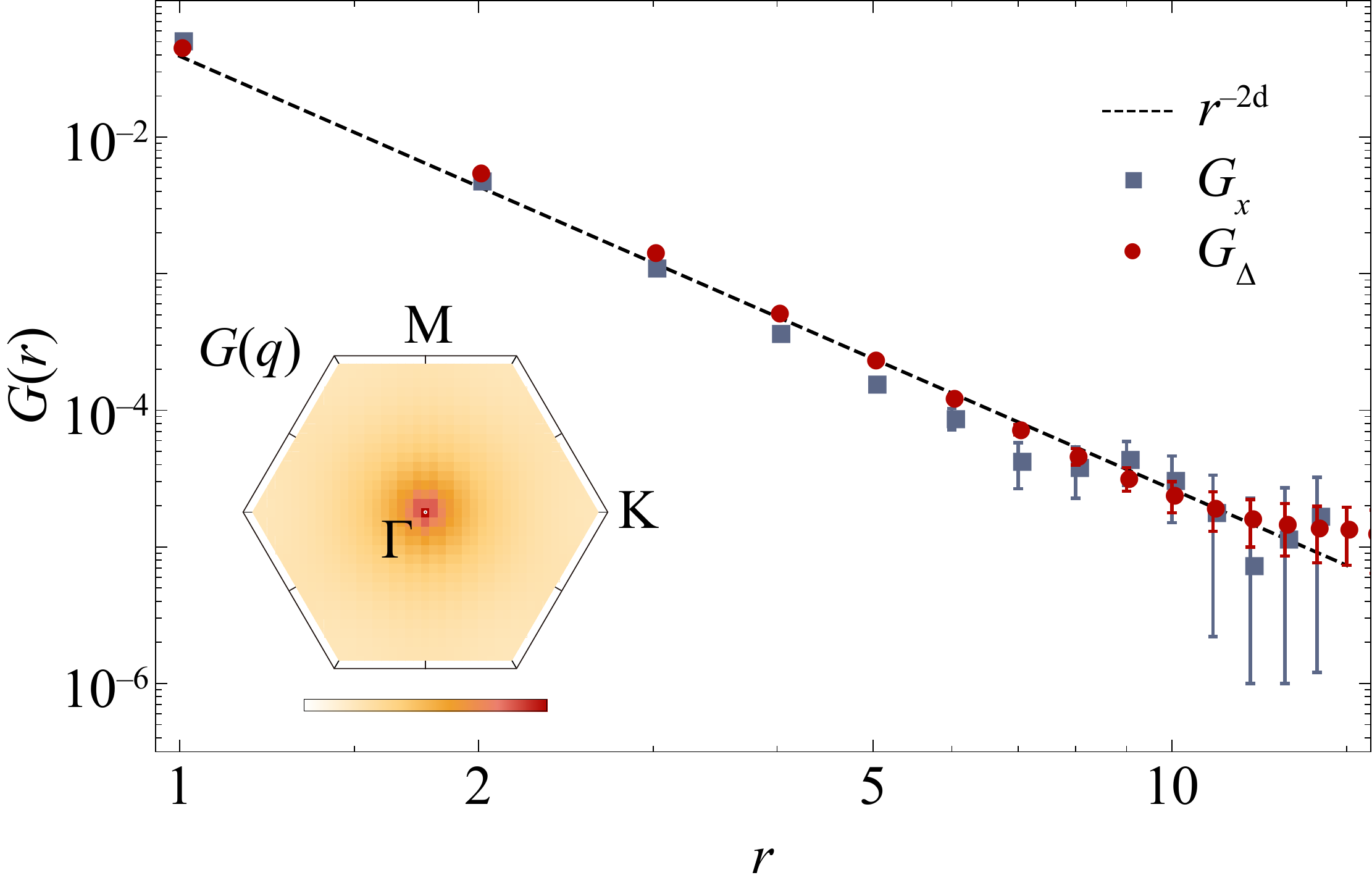}
    \caption{The scaling behavior characterized by UV-IR mixing. The correlation function of triangle spin $G_\triangledown(r)$ and $x$-spin $G_x(r)$ agrees with the theoretical prediction $G(r)\sim r^{-2d}$, measured with system size $L=64$ at critical point $\Gamma=1$. Inset: The structure factor $G_x(\mathbf{q})$ from the $G_x(r)$ correlator plotted in the first Brillouin Zone, measured with system size $L=31$ at critical point $\Gamma=1$. }
    \label{fig_4}
\end{figure} 

\textit{Fractal quantum criticality}---One elusive feature about this novel quantum criticality is the fractal scaling dimension. 
As shown in Fig.~\ref{fig_4}, the energy density correlation $G_x(r)=\langle\sigma^x(0)\sigma^x(r)\rangle$ and its dual $G_\triangledown(r)=\langle\tau^x(0)\tau^x(r)\rangle$ display a power-law decay $G_\triangledown(r)\sim r^{-2d}$ at the critical point, with $d=\ln(3)/\ln(2)$ being the fractal dimension. In the meantime, the manybody correlator $C(r)$ in Eq.~(\ref{eq_cor}) displays a short-ranged correlation at the phase transition point.
This implies that the charge operator $\sigma^x$ at the critical point carries a fractal dimension, albeit the lattice model we begin with is defined on a 2D lattice with local interactions. 
Usually, the scaling exponent of the energy density correlation $S(r)=r^{-D}$ depends on the space-time dimension $D$. 
Such universality at the critical point is based on the assumption that the short-wavelength mode at the phase transition point is irrelevant so that we can coarse-grain the local fluctuation and only keep the long wave-length modes in our effective low energy theory.

The characteristic fractal scaling behavior we obtain here represents a type of quantum criticality beyond the renormalization group paradigm. 
Due to the fractal symmetry in our model, the critical point contains numerous low-energy modes with rough field configurations originating from short-wavelength fluctuation. 
These modes with strong local fluctuation survive in the low energy spectrum at the critical point and the effective theory is mixed with UV degree of freedom~\cite{paramekanti2002ring,xu2007bond,you2021fractonic,you2020fracton,seiberg2020exotic,gorantla2021modified}. 
Thus the critical theory is affected by the short-wavelength physics and the standard renormalization group picture does not apply due to the existence of subextensive UV modes at low energy. 

Let us now provide an intuitive picture to explain the fractal scaling dimension. 
We begin with the Gaussian theory~\cite{bulmash2018generalized},
\begin{align}\label{ga}
    & \mathscr L=K(\partial_t \theta)^2+K (D_i \theta)^2\nonumber\\
    & D_i=a_0\nabla_x \nabla_y+1/a_0+\nabla_y.
\end{align}
Here $a_0$ is the lattice spacing, $\theta$ is an Ising variable as $e^{i \theta}=\sigma^z$, and $D_i$ is a lattice differential polynomial that creates the triangle coupling with three-body interaction.
The Green function $G(r,t)=\langle\theta(0,0)\theta(r,t)\rangle$ of this Gaussian theory satisfies the saddle point equation,
\begin{align}
    &K(\partial_t^2+ D_i^2)G(r,t)=\delta(r) \delta(t).
\end{align}
To get an approximate solution for this Green function, we first define $\int\mathrm dS_f$ as the congujate operator of $D_i$ ($[\int\mathrm dS_f, D_i]=1$) with $\mathrm dS_f$ being the integral over the area inside the Sierpinski triangle. As the fractal has dimensionality $d=\ln(3)/\ln(2)$, the fractal area integral scale with distance with the asymptotic form $\int\mathrm dS_f \sim  r^d$.
The asymptotic behavior of the Green function at the critical point is
\begin{align}
G(r,t)=\frac{1}{K}\ln\left(\sqrt{t^2+r^{2d}}\right).
\end{align}
Correspondingly, the spatial correlation function for $\sigma^x$ reads,
\begin{align}
G_x(r,t=0)=\langle\sigma^x(0)\sigma^x(r)\rangle=\langle\partial_t \theta(0)\partial_t \theta(r)\rangle\sim r^{-2d}
\end{align}
This agrees exactly with the numerical result shown in Fig.~\ref{fig_4}. 
The fractal scaling dimension is thus a direct sequence of the three-body coupling that respects the fractal symmetry. 
If we generate a rough field configuration of $\theta$ by implementing the fractal $\mathbb Z_2$ transformation to the Gaussian theory in Eq.~(\ref{ga}), the action is still invariant, such that the low energy modes at the critical point contain a large number of rough field configurations connected by the fractal symmetry. 
The survival of these rough patterns at the critical point engenders the ``UV-IR mixing'' as the low energy degrees of freedom are manipulated by the physics at short-wavelength such the traditional renormalization group paradigm does not apply. 
Such UV-IR mixing is a peculiar feature in fracton systems ascribable to the subsystem symmetry that enables the short-wavelength fluctuation to outlive at low energy. 

The UV-IR mixing also shows characteristic signatures in the structure factor $G_x(\mathbf{q})=\sum_{ij}\langle\sigma_i^x\sigma_j^x\rangle\mathrm e^{\mathrm i\mathbf q\cdot(\mathbf r_i\cdot\mathbf r_j)}$ as seen in the inset of Fig.~\ref{fig_4}. Besides, due to the excessive low-energy states at finite momentum, we expect the quantum critical point might display non-additive long-range entanglement entropy whose scaling exceeds the area-law. 
Unlike common quantum critical points, whose low energy excitation are ascribed by long wave-length modes and a minimum at the $\Gamma$ point, we observe an extensive region in momentum space with low intensity, corresponding to strong fluctuations of defects at short wave-length with low energy cost~\cite{paramekanti2002ring,xu2007bond, you2020fracton,bulmash2018generalized}.
On the other hand, the zero momentum has higher intensities since low momentum has finite energy~\cite{bulmash2018generalized}. 

\emph{Conclusion and outlook}--- We identified a continuous quantum phase transition between the paramagnetic phase and the fractal symmetry breaking phase of the quantum Newman-Moore model. 
A significant new element of this critical theory is that the infrared (IR) effective theory is controlled by short-wavelength fluctuation with peculiar UV-IR mixing. 
Our work has identified hallmarks of fractal symmetry at quantum critical points and opens new ground for future studies of phase transitions with fracton dynamics.
Moreover, due to the dynamical constraints of the quasi-particle excitations, the system exhibits glassy behavior and creates fertile ground to study quantum annealing problems with potential application to quantum memories. 

\emph{Acknowledgement}--- 
We are grateful for many useful exchanges with Cenke Xu.
This work was supported by the EuropeanResearch Council (ERC) under the European Unions Horizon2020 research and innovation program (grant agreement No.771537). 
F.P. acknowledges the support of the Deutsche Forschungsgemeinschaft (DFG, German Research Foundation) under Germany's Excellence Strategy EXC-2111-390814868 and TRR 80.

\clearpage

\setcounter{figure}{0}
\setcounter{section}{0}
\setcounter{equation}{0}
\renewcommand{\theequation}{S\arabic{equation}}
\renewcommand{\thefigure}{S\arabic{figure}}

\onecolumngrid

\newcommand{\vsigma}{\mbox{\boldmath $\sigma$}}

\section*{\Large{Supplemental Material}}

\section{Self-duality}

\begin{figure}[b]
    \centering
    \includegraphics[width=0.3\linewidth]{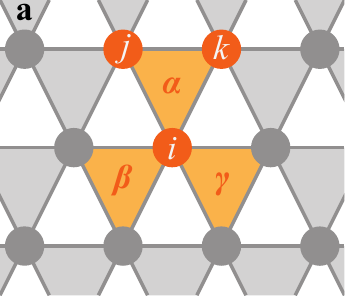}
    \includegraphics[width=0.4\linewidth]{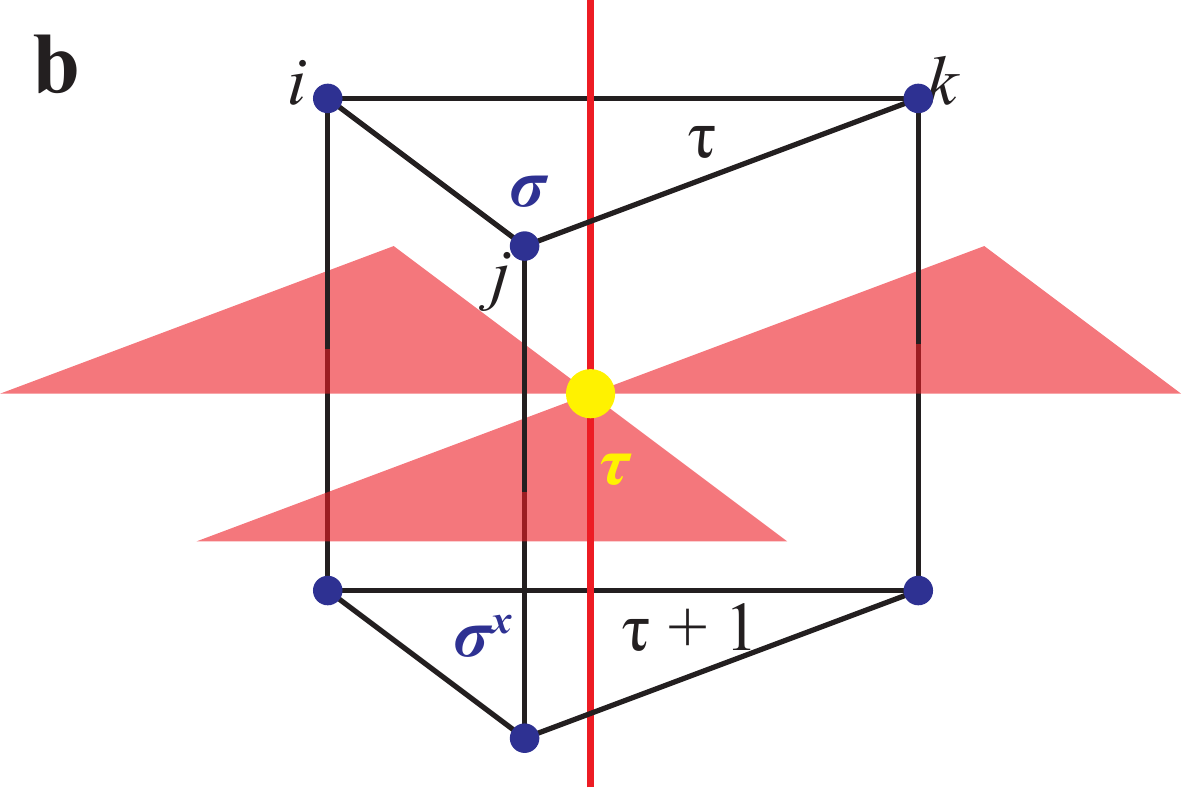}
    \caption{(a) An illustration of the lattice setup and the $\sigma$--$\tau$ self-duality. (b) An illustration of the mapping onto the 3D classicial model on the layered triangular lattice, the black and red lines denotes the triangular prism cage and the dual bonds, respectively. }
    \label{fig_s1}
\end{figure}

\begin{figure}[t]
    \centering
    \includegraphics[width=0.5\linewidth]{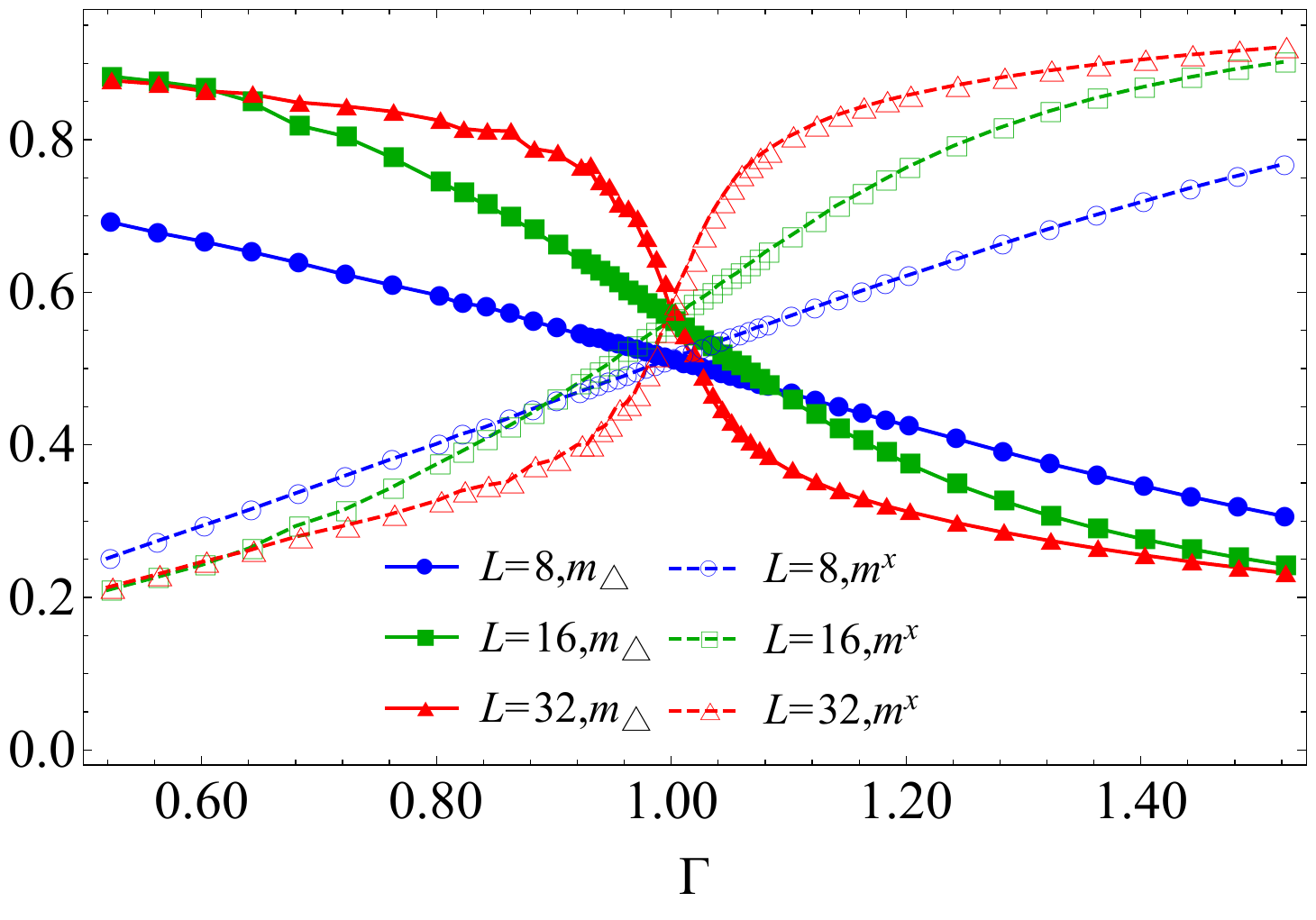}
    \caption{The polarizations $m_\triangle=\langle\tau^x\rangle$ and $m^x\langle\sigma^x\rangle$ in different system sizes.}
    \label{fig_s2}
\end{figure}

The quantum Newman-Moore model exhibits a Krammer-Wanier self-duality \cite{fracton_rev_2}. To elucidate this duality formulism, we define an effective spin-$1/2$ $\tau_\alpha$ at the center of each downward triangle on the dual lattice (Fig.~\ref{fig_s1}a) with
\begin{equation}
    \tau^x_\alpha=\sigma_i^z\sigma_j^z\sigma_k^z.
\end{equation} 
In this way, we can rewrite the dual-Hamiltonian as follows. The $\sigma_i^z\sigma_j^z\sigma_k^z$ term becomes the onsite Zeeman field $\tau^x_\alpha$ while the spin-flip term $\sigma^x$ is dual to the three-body interaction among the three spins on the upward triangle of the dual lattice $\sigma^x_i=\tau^z_\alpha \tau^z_\beta \tau^z_\gamma$,
\begin{equation}
    H=-J\sum_\alpha\tau^x_\alpha-\Gamma\sum_\triangle \tau^z_\alpha\tau^z_\beta\tau^z_\gamma.
\end{equation}
This is exactly the same as the original model with interchanging coupling constant $J$ and $\Gamma$. The self-duality can be verified numerically by measuring the polarizations $m_\triangle=\langle\tau^x\rangle$ and $m^x\langle\sigma^x\rangle$. We find that these two curves are symmetrical with respect to the $\Gamma=\Gamma_c$ vertical line in Fig.~\ref{fig_s2}. 

The self-duality is also manifested by mapping the path integral of this 2D quantum model to the partition function of a 3D classical model on the layered triangular lattice at finite temperature \cite{duality}
\begin{equation}
    S=-\tilde{J}\sum_{(ijk)_\triangledown,\tau}\sigma_{i\tau}^z\sigma_{j\tau}^z\sigma_{k\tau}^z-\tilde{\Gamma}\sum_{i,\tau}\sigma_{i\tau}^z\sigma_{i,\tau+1}^z,
\end{equation}
where $\tilde{J}=J\delta\tau$ and $\mathrm e^{-2\tilde{\Gamma}}=\tanh\delta\tau$, $\delta\tau$ being the imaginary time slice. The partition function of the above model is,
\begin{equation}
    Z=\sum_{\{\sigma\}}\mathrm e^{-S}=\sum_{\{\sigma\}}\prod_\triangledown\left(\cosh\tilde{J}+\sigma_1^\triangledown\sigma_2^\triangledown\sigma_3^\triangledown\sinh\tilde{J}\right)\times\prod_b\left(\cosh\tilde{\Gamma}+\sigma_1^b\sigma_2^b\sinh\tilde{\Gamma}\right) .
\end{equation}
We introduce triangle variables $k_\triangledown=0,1$ and bond variables $k_b=0,1$ and define $c_0=\cosh\tilde{J}$, $c_1=\sinh\tilde{J}$, $d_0=\cosh\tilde{\Gamma}$, $d_1=\sinh\tilde{\Gamma}$. The partition function can be interpreted as,
\begin{equation}
    Z=\sum_{\{s\}}\sum_{\{k_\triangledown,k_b\}}\prod_\triangledown c_{k_\triangledown}(\sigma_1^\triangledown\sigma_2^\triangledown\sigma_3^\triangledown)^{k_\triangledown}\times\prod_bd_{k_b}(\sigma_1^b\sigma_2^b)^{k_b} .
\end{equation}
First, take the spin sum: each spin is summed to two if raised to an even power and zero otherwise. This results in a constrained sum over $k$ variables:
\begin{equation}
    Z=2^N\sum'_{\{k_\triangledown,k_b\}}\prod c_{k_\triangledown}\prod d_{k_b}.
\end{equation}
Each site of the original lattice is connected to three downward triangles and two $z$-bonds with the constraint that the sum of the five $k$ variables is even for all sites. 

We introduce the dual spins $\tau$ located at the center of each cage-unit on the layered triangular lattice. For site $i$ on the original lattice, its three neighboring down-pointing triangles are pierced by vertical bonds of the dual lattice. For each piercing bond $b$ of the dual lattice, we set $k_\triangledown=(1-\tau^b_1\tau^b_2)/2$. Similarly, each of the two vertical bonds $b$
containing site $i$ pierces a down-pointing triangle of the dual lattice, and we set $k_b=(1-\tau_1^\triangledown\tau_2^\triangledown\tau_3^\triangledown)$. The $k$ variables given by $\tau$ dual spins automatically satisfy the constraints.

Now we calculate the dual couplings, 
\begin{equation}
    \begin{aligned}
        c_{k_\triangledown}&=k_\triangledown\sinh\tilde{J}+(1-k_\triangledown)\cosh\tilde{J}\\
        &=\frac{1+\tau_1^b\tau_2^b}{2}\cosh\tilde{J}+\frac{1-\tau_1^b\tau_2^b}{2}\sinh\tilde{J}\\
        &=\frac{\mathrm e^{\tilde{K}}}{2}(1+\tau_1^b\tau_2^b\cosh\tilde{\Gamma}^*)\\
        &=(2\sinh(2\tilde{\Gamma}^*))^{-1/2}\mathrm e^{\tilde{\Gamma}^*\tau_1^b\tau_2^b},
    \end{aligned}
\end{equation}
where we have defined $\tanh\tilde{\Gamma}^*=\mathrm e^{-2\tilde{J}}$. 

By the same process 
\begin{equation}
    d_{k_b}=(2\sinh(2\tilde{K}^*))^{-1/2}\mathrm e^{\tilde{K}^*\tau_1^\triangledown\tau_2^\triangledown\tau_3^\triangledown}
\end{equation}
with $\tanh\tilde{K}^*=\mathrm e^{-2\tilde{\Gamma}}$. These relations determine the self-dual point at $\tilde{J}=\tilde{\Gamma}=\log\frac{1+\sqrt{2}}{2}$. 

\section{Fractal symmetry and polynomial representation}

When periodic boundary conditions are imposed, the ground state degeneracy oscillates with the system size. The symmetry operator which connects these degenerate ground state is known to be a ``subsystem symmetry''. A subsystem symmetry operation acts only on a subset of the lattice sites with a subextensive degree of freedom. Unlike global symmetries which act on the whole system, here the subsystem symmetry acts on a fractal part of the lattice with the shape of a Sierpinski-triangle.

A more elegant formulation of such subsystem symmetries is provided in terms of a polynomial representation. We give here a minimal description and refer to Ref.~\cite{fracton_rev_2} for more details. Every site $(i,j)$ is represented by a term $x^iy^j$. The factors of the terms are $\mathbb{Z}_2$ numbers which take value only from $\{0,1\}$ and $1+1=0$. The Sierpinski triangle, for example, can be expressed as
\begin{equation}
    F(x,y)=\sum_{k=0}^{\infty}(y+xy)^k.
\end{equation}
The terms in the Hamiltonian $\sigma^z_{ij}\sigma^z_{i,j-1}\sigma^z_{i-1,j-1}$ can be written as $\mathscr{Z}(x^iy^j(1+y^{-1}+x^{-1}y^{-1}))$, where $\mathscr{Z}$ maps a polynomial into the product of a set of Pauli-$z$ matrices, 
\begin{equation}
    \mathscr{Z}\left(\sum_{ij}c_{ij}x^iy^j\right)=\prod_{ij}(\sigma_{ij}^z)^{c_{ij}}
\end{equation}
and $\mathscr{X}$ is similarly defined. In this way we can express the Hamiltonian as
\begin{equation}
    H=-J\sum_{ij}\mathscr{Z}(x^iy^j(1+y^{-1}+x^{-1}y^{-1}))-h\sum_{ij}\mathscr{X}(x^iy^j)
\end{equation}
and the symmetry operator can be written as 
\begin{equation}
    S=\mathscr{X}(q(x)F(x,y)).
\end{equation}
For a half infinite plane, the $q(x)$ is an arbitrary polynomial; for a torus $q(x)$ has to satisfy an additional condition 
\begin{equation}
    q(x)(1+x)^L=q(x)
\end{equation}
with $x^L=1$. This condition breaks down into a system of linear equations over $\mathbb{Z}_2$, from which we can obtain the degeneracy,
\begin{equation}
    \log_2N(L)=\left\{\begin{array}{ll}
        2^m(2^n-2),&L=C\times 2^m(2^n-1)\\
        0,&\textrm{otherwise.}
    \end{array}\right. 
\end{equation}

\section{Many-body correlation and order parameter}

\begin{figure}
    \centering
    \includegraphics[width=0.5\linewidth]{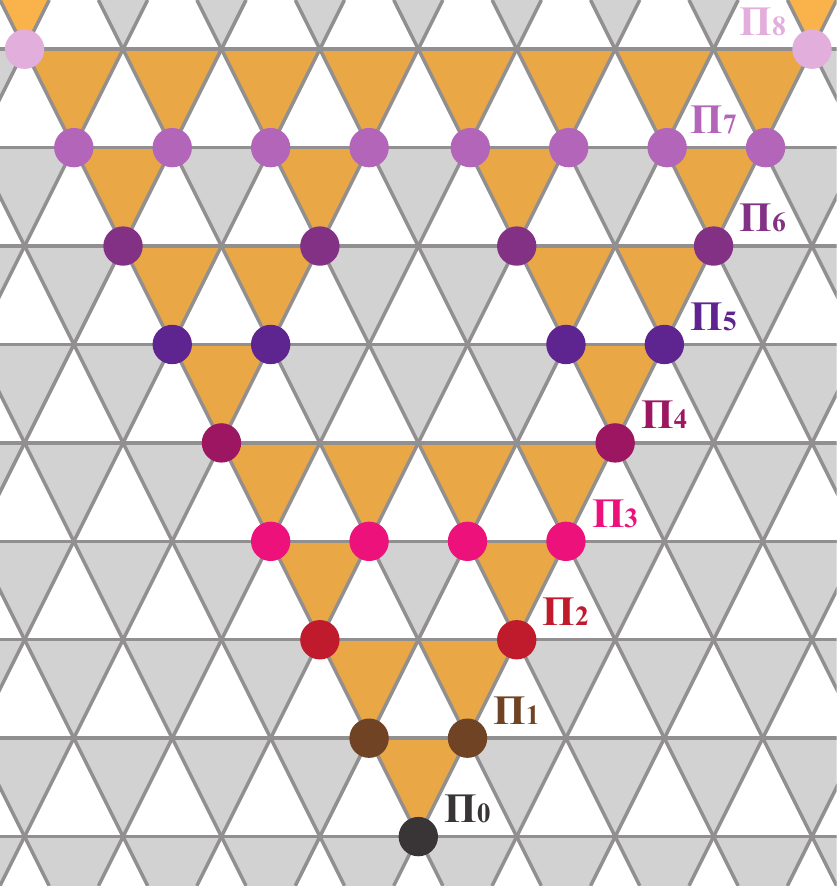}
    \caption{An illustration of the $\Pi_i$ operator as a generator of the many-body correlation, where the points of different color denotes the $\sigma^z$'s multiplied in the $\Pi_i$ operator.}
    \label{fig_s3}
\end{figure}

The many-body correlation operator can characterize the quantum phase transition between fractal symmetry breaking phase and paramagnetic phase~\cite{duality,fracton_rev_2}
\begin{equation}
    C(r)=\mathscr{Z}(1+(y^{-1}+x^{-1}y^{-1})^r).
\end{equation}
This polynomial representation of the correlator can be interpreted as the product of the spins on the corner together with all the spins in the $r$-th line of a down-pointing Sierpinski triangle as shown in Fig.~\ref{fig_s3}, \textit{e.g.},  $C(r=3)=\sigma_{0,0}^z\sigma_{\bar 3,0}^z\sigma_{\bar 3,\bar 1}^z\sigma_{\bar 3,\bar 2}^z\sigma_{\bar 3,\bar 3}^z$ and $C(r=4)=\sigma_{0,0}^z\sigma_{\bar 4,0}^z\sigma_{\bar 4,\bar 4}^z$, where the bar denotes a negative sign. This operator reduces to a three-body correlation when $r=2^k$. The operator is expected to be non-vanishing in the fractal symmetry breaking phase and decays hyper-exponentially as $C(r)\sim\mathrm e^{-(r/\xi)^d}$ \cite{fracton_rev_2} in the paramagnetic phase, where $d=\ln 3/\ln 2$ is the fractal dimension. 
    
    The nonvanishing expectation value of $C(r)$ in the symmetry-breaking phase implies the quantum correlation among the spins living at the corners of the Sierpinski triangle. It is noteworthy to mention that the $C(r)$ depends on the number $N_r$ of $\sigma_z$ in the product, which can be regarded as the number of ``corners'' on the fractal boundary. Such fractal scaling can be visualized from the dual picture.
 As $C(r)$ is the product of all $\tau_x=\sigma_i^z\sigma_j^z\sigma_k^z$  operators inside the Sierpinski fractal, it counts the total number of dual $Z_2$ charges ($\tau_x$) inside the fractal region. The transverse field $\sigma_i^x$ creates three fracton defects with $\tau_x=\sigma_i^z\sigma_j^z\sigma_k^z=-1$ in the adjacent downward triangle. In the fractal symmetry breaking phase, such fracton defect is confined and forms a
  three-fracton bound state so the defect is always created in triads as Fig.~\ref{fig_s1}. These triple defect excitations could
reverse the dual-charge inside the Sierpinski triangle if and only if they live at the corner of the fractal. Thus, the scaling of many-body operator $C(r)$ obeys a ``corner-law''. Likewise, in the paramagnetic phase, the isolated defects proliferate and immerse into the fractal. Each isolated defect with $\tau_x=-1$ inside the fractal flips the parity of the dual charges so the expectation value of the dual charge inside the fractal has a hyper-exponentially scaling $C(r)\sim\mathrm e^{-(r/\xi)^d}$ with $r^d$ being the number of sites inside the fractal.

\begin{figure}[ht]
    \centering
    \includegraphics[width=0.5\linewidth]{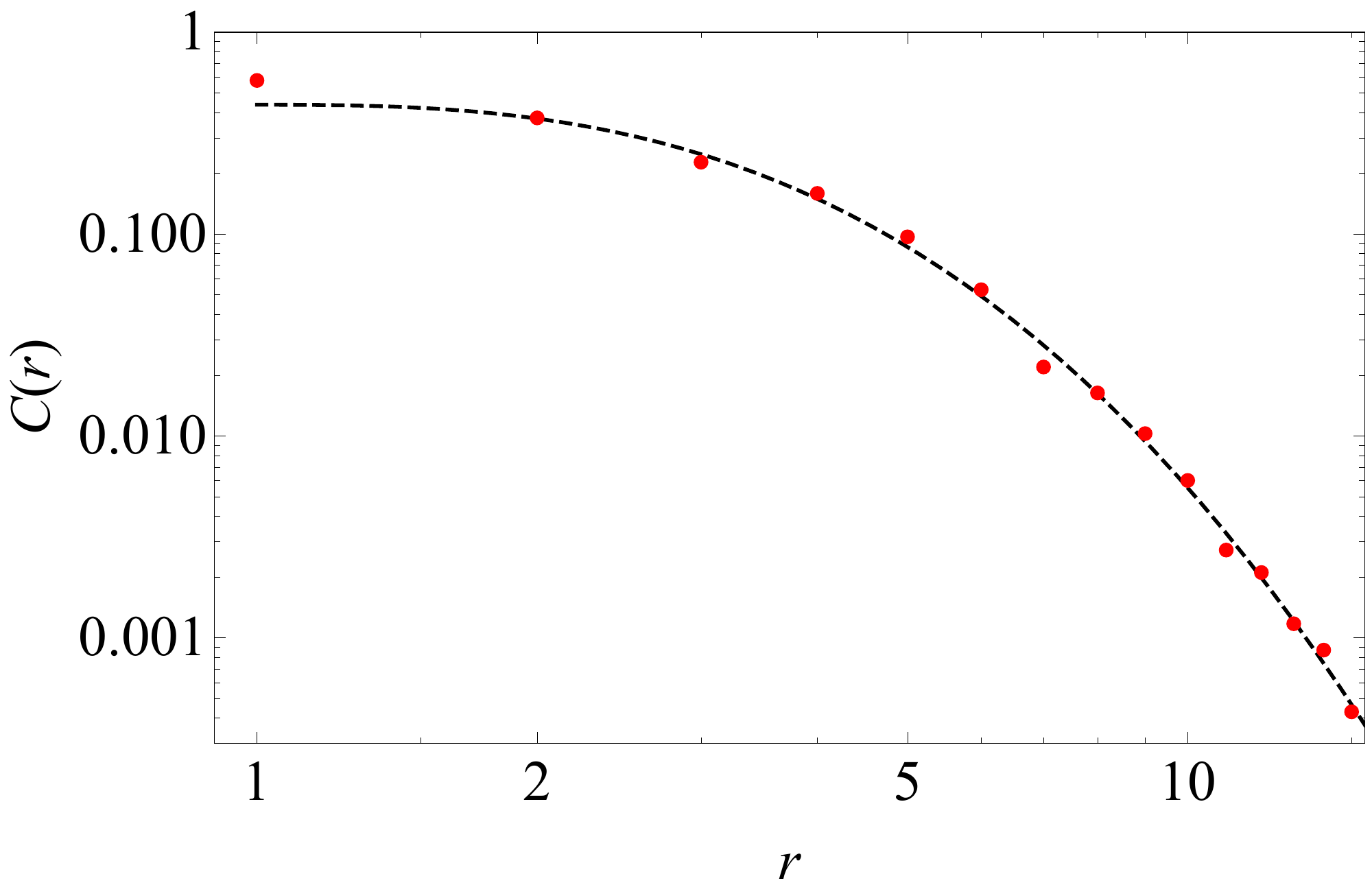}
    \caption{The many-body correlation function $C(r)$ at the quantum critical point plotted on a log–log graph; red points --- Monte Carlo results; black dashed line --- fitting in the form $C(r)=C_0e^{-C_1(\ln(r))^a}$.}
    \label{qcpfig}
\end{figure}

At the quantum critical point, the many-body operator $C(r)$ displays short-ranged correlation. In particular, $C(r)$ decreases with distance $r$ in a manner that is faster than any power-law function but slower than the exponential function. In Fig.~\ref{qcpfig}, we plot the correlation function in the log–log graph and its finite scaling implies $C(r)\sim e^{-C_1(\ln(r))^a}$ with $a=2.7(1)$. 

We can generalize the expression by introducing another subscript
\begin{equation}
    C_{rl}=\mathscr{Z}((y^{-1}+x^{-1}y^{-1})^r+(y^{-1}+x^{-1}y^{-1})^l).
\end{equation}
Taking the summation over $r$ and $l$, 
\begin{equation}
    \sum_{rl}C_{rl}=\sum_{rl}\mathscr{Z}((y^{-1}+x^{-1}y^{-1})^r)\mathscr{Z}((y^{-1}+x^{-1}y^{-1})^l)=L^2|\psi|^2
\end{equation}
the summation can be decomposed into the square of an ``order parameter'', defined as 
\begin{equation}
    \psi=\frac{1}{L}\sum_{r=0}^{L-1}\mathscr{Z}((y^{-1}+x^{-1}y^{-1})^r).
\end{equation}
This quantity genuinely distinguishes the ordered phase and the paramagnetic phase.

\section{UV-IR mixing at criticality --- How renormalization breakdown}
\begin{figure}[t]
    \centering
    \includegraphics[width=0.35\linewidth]{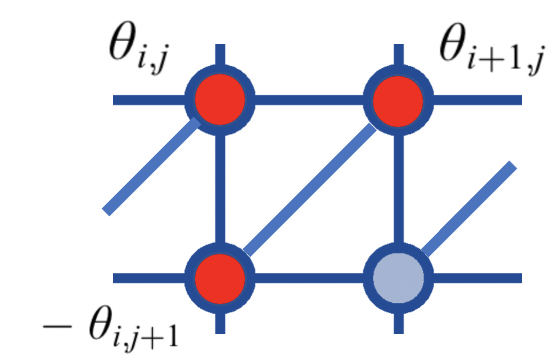}
    \caption{The three-body interaction on a triangle can be expressed as $\cos(\theta_{i,j}+\theta_{i+1,j}-\theta_{i,j+1})$.}
    \label{th}
\end{figure}

In the manuscript, we build a Gaussian theory to explain the fractal scaling dimension at the phase transition point
\begin{align}\label{ga}
    & \mathscr L=K(\partial_t \theta)^2+K (D_i \theta)^2\nonumber\\
    & D_i=a_0\nabla_x \nabla_y+1/a_0+\nabla_y.
\end{align}
Here $a_0$ is the lattice spacing and $D_i$ is a lattice differential polynomimal that creates the three-body interaction. We illustrate this differential polynomimal formulism in Fig.~\ref{th} on the tilted triangular lattice. Each site contains an Ising degree of freedom $\sigma^z=e^{i\theta},\sigma^x=e^{i \pi n}$. Here $\theta,n$ are conjugate pairs with discrete values $\theta=0,\pi$ and $n=0,1$.
The three-body coupling can be written as
\begin{align}
\cos(\theta_{i,j}+\theta_{i+1,j}-\theta_{i,j+1}).
\end{align}
Expand the cosine term to the quadratic level and replace lattice difference with differentials,
\begin{align}
&(\theta_{i,j}+\theta_{i+1,j}-\theta_{i,j+1})^2\nonumber\\
&=[(\theta_{i,j}-\theta_{i+1,j}-\theta_{i,j+1}+\theta_{i+1,j+1})+(\theta_{i+1,j}-\theta_{i+1,j+1})+\theta_{i+1,j}]^2\nonumber\\
&=((a_0\nabla_x \nabla_y+\nabla_y+1/a_0) \theta)^2\equiv (D_i\theta)^2.
\end{align}
The Newman-Moore model becomes,
\begin{align}
&\mathcal{L}=n \partial_t \theta+\Gamma (D_i \theta)^2+J n^2.
\end{align}
Integrating out the density fluctuation $n$, we get
\begin{align}
&\mathcal{L}=\frac{1}{4J} (\partial_t \theta)^2+\Gamma (D_i \theta)^2.
\end{align}
Here we keep the value of $4J/\Gamma$ fixed and tune the strength of $\Gamma$ throughout the phase transition. After a rescaling of time, we get
\begin{align}
&\mathcal{L}=K (\partial_t \theta)^2+K (D_i \theta)^2.
\end{align}
Notably, this differential polynomial $D_i$ contains a constant piece which implies the global charge conservation is broken due to the three-body interaction while the fractal symmetry is still respected. 

At the critical point, the gapless excitation has a dispersion $\epsilon(k)=|a_0 k_x k_y+1/a_0+k_y|$. Here the lattice regularization $a_0\sim 1/\Lambda$ is essential due to subsystem symmetry and we will keep the momentum cutoff as our UV regulator. Peculiarly, the energy minimal does not appear at the $\Gamma$ point. This can be manifested from the absence of global $Z_2$ symmetry. In addition, there should exist a large region at finite momentum with zero energy. This agrees with our numerical result for the static structure factor
\begin{align}
G_x(\mathbf{q})=\int d\omega ~~\omega^2 G(q,\omega) 
=\int d\omega ~~\frac {\omega^2}{K(\omega^2+\epsilon^2(q))}=\epsilon^2(q)/K.
\end{align}
Unlike the other critical phenomenon whose low energy excitations are ascribed by long wave-length modes and the structure factor minimum appears at the $\Gamma$ point, the structure factor $G_x(\mathbf{q})$ at the fractal symmetric critical point contains an extensive region in the momentum space with low intensity, corresponding to the strong fluctuation of defects at short wave-length. On the other hand, the zero momentum had a higher intensity since zero momentum modes cost finite energy. 
Therefore the low-energy dynamics at the phase transition does not occur at long wave-length region and the short-wavelength physics controls the critical behavior.

The quantum critical points connecting different phase patterns
have broad indications and deep impact from various aspects. The divergence of the correlation length at the critical region implies the effective interaction at IR becomes highly non-local and hence requires us to visualize the system at a larger scaling at long wavelength. A complete understanding of critical phenomena is accomplished via the development of a renormalization group theory. The RG theory of critical phenomena provides the natural framework for defining quantum field theories at a nonperturbative level and hence has broad implications for strong coupling problems. In particular, the universal properties of a broad class of random or statistical systems can be understood by coarse-graining the short-wavelength physics with local fluctuation and focus on the long-wavelength behaviors. Based on this observation, the critical phenomenon shares many universal properties that are independent of the UV Hamiltonian but only rely on symmetry and dimensionality. For instance, the scaling dimension and the correlation function with respect to the energy density at the critical point (phase) have a universal power-law exponent $S(r)=r^{-D}$ depends on the space-time dimension $D$.  Such a universal power law lies in the fact that the low energy part of the spectrum is contributed by long-wavelength physics fluctuations at small momentum. Based on this assumption, the field patterns at low energy, which control the IR behavior, are determined by the spatial dimension and dynamical exponent that are insensitive to UV cut-offs.

However, the fractal symmetry breaking phase transition we demonstrate here is a peculiar example that escapes from the renormalization paradigm. The essence of such RG breakdown lies in the fact that the subsystem symmetry in the many-body system engenders a large number of collective excitations with strong fluctuation at short wavelength but still survive in the low energy spectrum when proximate to the phase transition point. The fractal scaling dimension of the energy density correlation and the subextensive number of zero-energy modes at finite momentum is a direct manifestation of such UV-IR mixing.

In contrast to the spontaneous breaking of global symmetries, our phase transition theory is dominated by short-wavelength physics with local fluctuation. The low energy modes at the critical point contain a large number of rough field configurations connected by fractal symmetry. The survival of these rough patterns at the critical point engenders the ``UV-IR mixing'' as the low energy degree of freedom is manipulated by the physics at short wavelength and the traditional renormalization group paradigm does not apply. In particular, we cannot simply coarse-grain the local fluctuation nor change the UV cut-off as the high momentum modes with zero energy can bring additional singularity and hence qualitatively change the universal behavior.

\section{Numerical method}

For the Monte Carlo simulations, we use the stochastic series expansion (SSE) method. We adapt the transverse field Ising model update strategy \cite{qmc_2} for the three-body Ising operator. 

In quantum statistics, the measurement of observables is closely related to the calculation of partition function $Z$
\begin{equation}
    \langle\mathscr{O}\rangle=\mathrm{tr}\:\left(\mathscr{O}\exp(-\beta H)\right)/Z,\quad Z=\mathrm{tr}\exp(-\beta H),
\end{equation}
where $\beta=1/T$ is the inverse temperature, $H$ is the Hamiltonian of the system and $\mathscr{O}$ is an arbitrary observable. Typically, in order to evaluate the ground state property, one takes a sufficiently large $\beta$ such that $\beta\sim L^z$, where $L$ is the system scale and $z$ is the dynamical exponent. In SSE, such evaluation of $Z$ is done by a Taylor expansion of the exponential and the trace is taken by summing over a complete set of suitably chosen basis 
\begin{equation}
    Z=\sum_\alpha\sum_{n=0}^\infty\frac{\beta^n}{n!}\langle\alpha|(-H)^n|\alpha\rangle.
\end{equation}
We then write the Hamiltonian as the sum of a set of operators whose matrix elements are easy to calculate
\begin{equation}
    H=-\sum_iH_i.
\end{equation}
In practice, we truncate the Taylor expansion at a sufficiently large cutoff $M$ and it is convenient to fix the sequence length by introducing in identity operator $H_0=1$ to fill in all the empty positions despite it is not part of the Hamiltonian 
\begin{equation}
    (-H)^n=\sum_{\{i_p\}}\prod_{p=1}^nH_{i_p}=\sum_{\{i_p\}}\frac{(M-n)!n!}{M!}\prod_{p=1}^nH_{i_p}
\end{equation}
and 
\begin{equation}
    Z=\sum_\alpha\sum_{\{i_p\}}\beta^n\frac{(M-n)!}{M!}\langle\alpha|\prod_{p=1}^nH_{i_p}|\alpha\rangle.
\end{equation}

To carry out the summation, a Monte Carlo procedure can be used to sample the operator sequence $\{i_p\}$ and the trial state $\alpha$ with according to their relative weight
\begin{equation}
    W(\alpha,\{i_p\})=\beta^n\frac{(M-n)!}{M!}\langle\alpha|\prod_{p=1}^nH_{i_p}|\alpha\rangle.
\end{equation}
For the sampling, we adopt a Metropolis algorithm where the configuration of one step is generated based on updating the configuration of the former step and the update is accepted at a probability
\begin{equation}
    P(\alpha,\{i_p\}\rightarrow\alpha',\{i'_p\})=\min\left(1,\frac{W(\alpha',\{i'_p\})}{W(\alpha,\{i_p\})}\right).
\end{equation}
Diagonal updates, where diagonal operators are inserted into and removed from the operator sequence, and cluster update, where diagonal and off-diagonal operates convert into each other, are adopted in the update strategy. 

For the transverse field Ising model $H=J\sum_bS_{i_b}^zS_{j_b}^z-h\sum_i\sigma_i^x$, we write the Hamiltonian as the sum of the following operators
\begin{equation}
    \begin{aligned}
        H_0&=1\\
        H_i&=h(S_i^++S_i^-)/2\\
        H_{i+n}&=h/2\\
        H_{b+2n}&=J(1/4-S_{i_b}^zS_{j_b}^z),
    \end{aligned}
\end{equation}
where a constant is added into the Hamiltonian for convenience. For the non-local opdate, a branching cluster update strategy is constructed, where a cluster is formed in $(D+1)$-dimensions by grouping spins and operators together. Each cluster terminates on site operators and includes bond operators. All the spins in each cluster are flipped together at a probability $1/2$ after all clusters are identified. 

\begin{figure}
    \centering
    \includegraphics[width=0.3\linewidth]{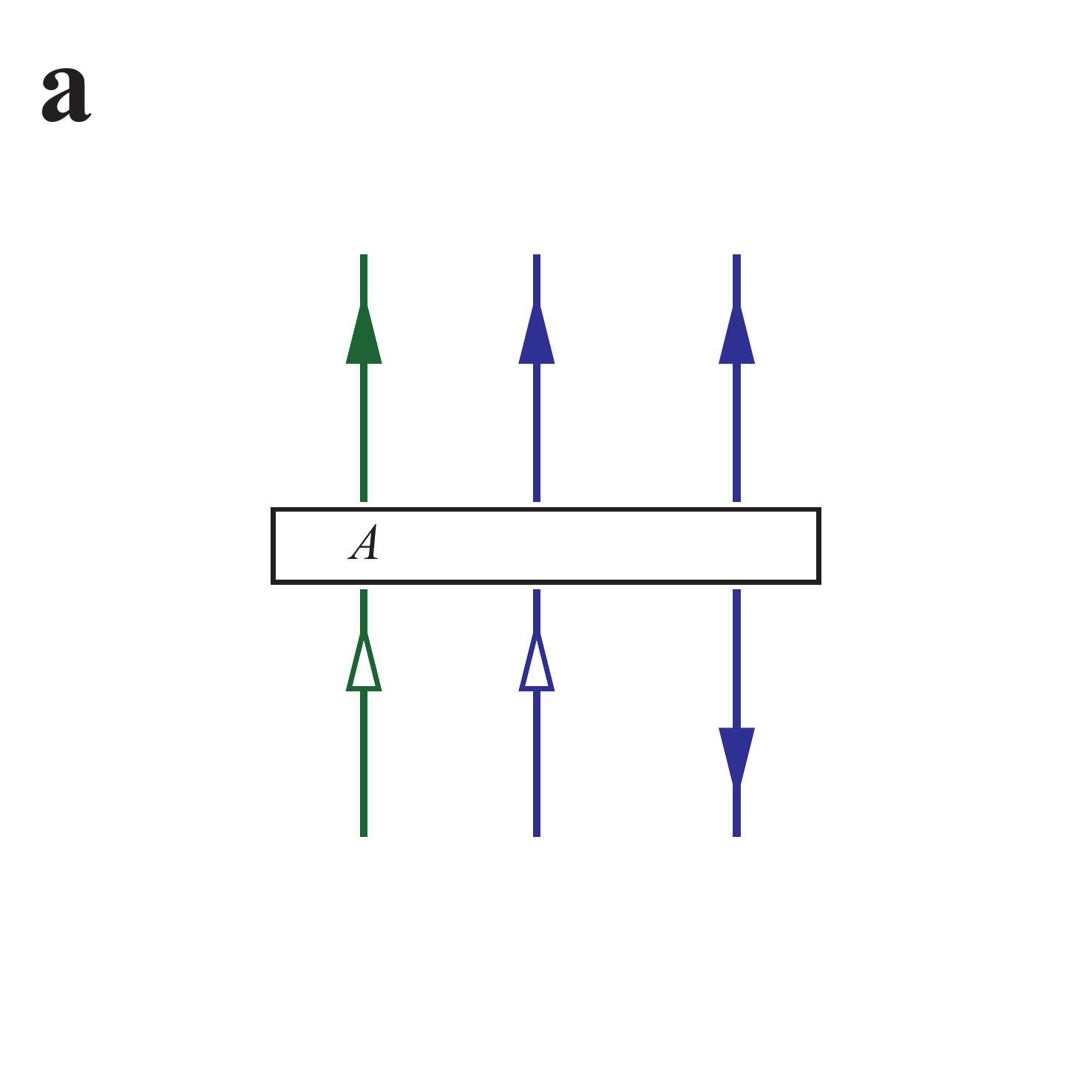}
    \includegraphics[width=0.5\linewidth]{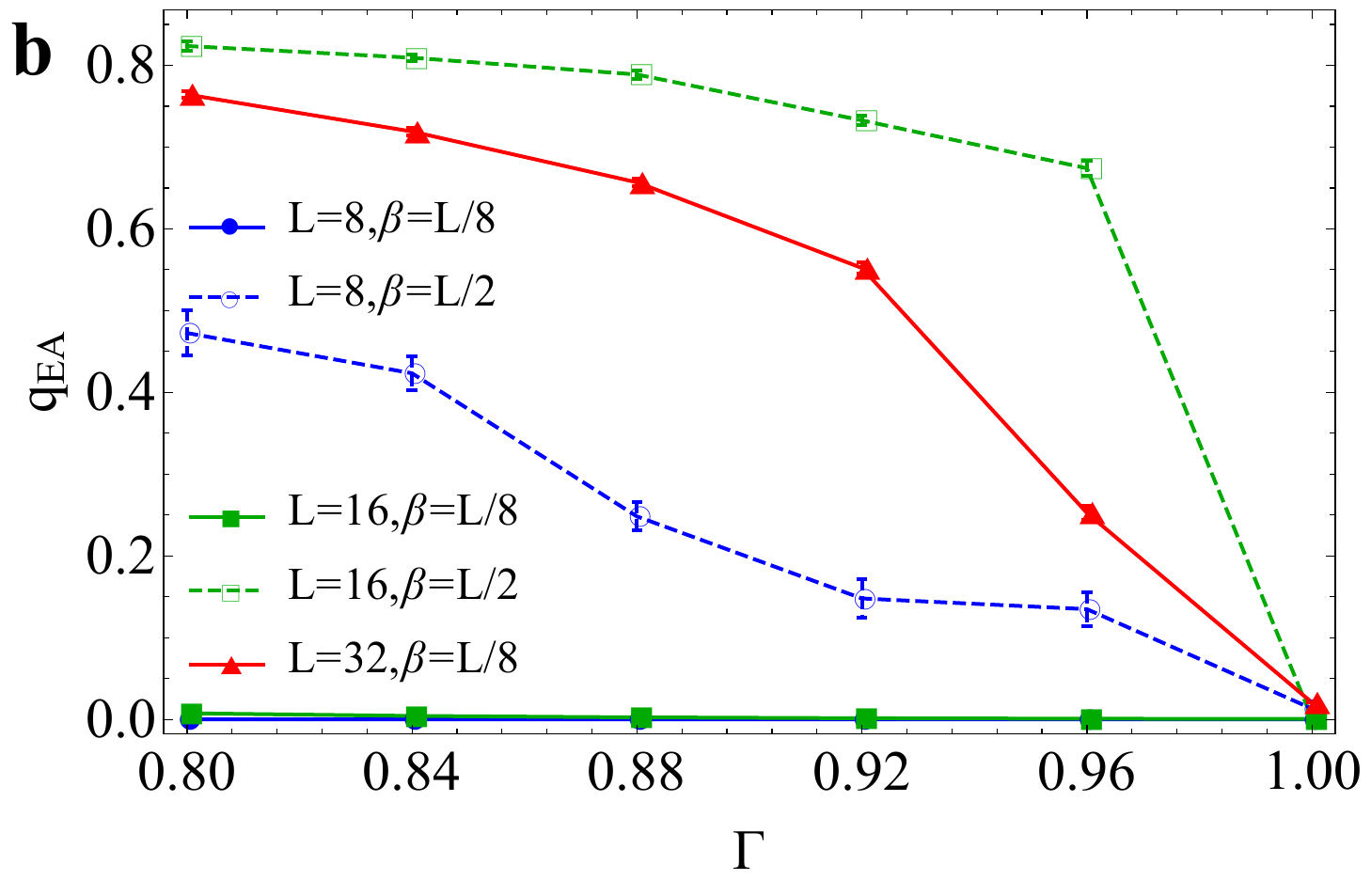}
    \caption[An illustration of the algorithm.]{(a) An illustration of our update strategy. (b) The Edwards-Anderson order parameter as a function of transverse field $\Gamma$ for different system sizes and temperatures. }
    \label{fig_s5}
\end{figure}

In our transverse field Newman-Moore model, we give values also to the triangular operators on which the constraint is broken
\begin{equation}
    \langle\uparrow\uparrow\uparrow|H_i|\uparrow\uparrow\uparrow\rangle=J+\epsilon, \langle\downarrow\downarrow\downarrow|H_i|\downarrow\downarrow\downarrow\rangle=\epsilon.
\end{equation}
On each triangular operator, one special site is chosen (denoted as $A$, the other two sites as $B,C$). When an update line enters site $B,C$, it exits from the other three legs at site $B,C$; when an update line enters site $A$, it only exits from the other \textit{one} leg at site $A$ (Fig. \ref{fig_s5}a). This update process will change the weight of the configuration, so each cluster is flipped at a probability $P_\mathrm{acc}=(J/\epsilon+1)^{\Delta N}$ calculated after the cluster is formed, where $\Delta N$ denotes the change in the number of energetically favorable operators.

\section{Glassiness and verification of ergodity}

Given the restricted mobility of excitations, the Monte Carlo simulation in such a system is easy to encounter the problem of glassiness \cite{newman_2}, which may undermine the ergodicity of our algorithm. To evaluate the glassiness, we calculate the Edwards-Anderson order parameter \cite{glass_order}
\begin{equation}
    q_\mathrm{EA}=\frac{1}{N}\sum_i\left(\langle S_i^z\rangle\right)^2.
\end{equation}
The limits $q_\mathrm{EA}=0$ and $1$ signify that the spins are completely flippable and completely frozen. We test three sizes $L=8,16$ and $32$ with different temperature $\beta=L/2$ and $L/8$ (Fig. \ref{fig_s5}b). When $\beta=L/8$, the system shows no glassiness for small sizes $L=8,16$. For $L=32$, the ergodicity of Monte Carlo updates seems acceptable in the vicinity of the transition point $\Gamma=1$, while glassiness shows up deep in the confined phase. When $\beta=L$ and $L=16$, the glassiness reaches a high level once we enter the confined phase. So we confine our following calculations to $L=8\beta$ without special indication. 

In order to reach the ground state in the ordered phase in spite of the glassiness, we adopt an annealing process by adding a longitudinal field
\begin{equation}
    H'=-h\sum_i\sigma^z_i.
\end{equation}
The field is lowered to zero during the thermalization process to restore the original Hamiltonian. 

\section{Finite size scaling}

To determine the critical exponents, we perform a finite-size scaling. It is worth noting that the order parameter we define is non-local, so that the exponents can only be formally defined. Their true significance remains to be studied. We choose 8 sizes between $8$ and $32$ at which the ground state is unique $L=8,11,13,16,20$, $23,26,32$ to perform the finite-size scaling.

The scaling hypothesis says that for any quantity $Q$, its finite size behavior should follow
\begin{equation}
    x_L(\Gamma)=(\Gamma/\Gamma_c-1)L^{1/\nu},\ y_L(\Gamma)=Q_L(\Gamma)L^{-\kappa/\nu},
\end{equation}
where $y$ is a universal function of $x$, regardless of the system size. The $\kappa$ denotes the exponent related to this quantity $Q$. \textit{E.g.}, for heat capacity $C$, $\kappa=\alpha$; for order parameter $\psi$, $\kappa=\beta$; for susceptibility $\chi$, $\kappa=\gamma$; for the dimensionless Binder ratio, we just have $\kappa=0$. As large sizes aren't reachable for us, correction terms are needed in some cases
\begin{equation}
    x_L(\Gamma)=(\Gamma/\Gamma_c-1)L^{1/\nu},\ y_L(\Gamma)=Q_L(\Gamma)\frac{L^{-\kappa/\nu}}{1+cL^{-\omega}}.
\end{equation}
To carry out the analysis in practice, we only retain the first terms in the Taylor expansion
\begin{equation}
    y=C_0+C_1x+C_2x^2+C_3x^3.
\end{equation}

The finite size scaling for the Binder ratio $U_\psi$ gives $\nu=0.50(2)$; the order parameter $\psi$ gives $\beta/\nu=0.448(8)$; the susceptibility gives $\gamma/\nu=2.19(4)$ and the heat capacity $C$ gives $\alpha/\nu=0.58(3)$. 

With the scaling relations, we can obtain the other exponents
\begin{equation}
    \gamma/\nu=2-\eta,
\end{equation}
which gives us a negative anomalous dimension $\eta=-0.19(4)$. 
All of our results on critical exponents are put together in Table \ref{tbl_1}.

\begin{table}[t]
    \caption{All the critical exponents calculated. }
    \begin{tabular}{c|c}
        \hline
        Exponent&Value\\
        \hline
        $\nu$&$0.50(2)$\\
        $\beta$&$0.224(5)$\\
        $\gamma$&$1.10(5)$\\
        $\alpha$&$0.28(2)$\\
        \hline
        $\eta$&$-0.19(4)$\\
        \hline
    \end{tabular}
    \label{tbl_1}
\end{table}

\end{document}